# SPACE-FILLING CURVES FOR QUANTUM CONTROL PARAMETERS

**Fariel Shafee**
**Department of Physics**

**Princeton University**
**Princeton, NJ 08544**

**Abstract:**

We consider the use of space-filling curves (SFC) in scanning control parameters for quantum chemical systems. First we show that a formally exact SFC must be singular in the control parameters, but a finite discrete generalization can be used with no problem. We then make general observations about the relevance of SFCs in preference to linear scans of the parameters. Finally we present a simple magnetic field example relevant in NMR and show from the calculated autocorrelations that a SFC Peano-Hilbert curve gives a smoother sequence than a linear scan.

## I. INTRODUCTION

A quantum system is represented by a state vector in Hilbert space, and the system has a Hamiltonian operator that determines how the system evolves by itself. The system may be in the ground state, in which case the evolution is simply a phase change with time of the state vector. If it is in an excited state then sooner or later it will drop to a lower state with either simply an emission of radiation, or more fundamental changes in the system. In a second quantized theory there is provision for considering annihilation and creation of particles; in a first quantized theory the transition of the dynamical system in terms of changes in its Hamiltonian description is more complicated. Feynman's path integral method is second quantized and yet it avoids explicit use of creation and annihilation operators. The transition in that case is implemented by a similar Hamiltonian as the second quantized case, and though an obvert operatorial interpretation is avoided the path integral actually achieves the same purpose of a discrete change in time from one set of fields to another. In particle physics, at relativistic energies this is a simple procedure.

In quantum chemistry usually we do not worry about changes in the nuclei, and hence with particle creation or destruction. The system is described by fixed nuclei and a number of electrons, all with the constant masses and charges. It is fairly easy to set up a Schrodinger equation, though not to solve it. If a large molecule dissociates, or if two molecules synthesize into a bigger one, no new particles are created or destroyed, they all correspond to the same Hamiltonian with the same nuclei and electrons as the components. However, the states vary. Like ionization being the free state of an electron, dissociation is also a new state of the system for the same Hamiltonian.

But it is possible to change the environmental factors of a system. That can be discrete or continuous. We introduce a control Hamiltonian that changes the overall development of the system. Populations can be inverted from the original ground state to some other desired state. Or other properties of the original state may be changed, e.g. expectation values of an observable. If this control Hamiltonian is a function of two variables, and these variables are continuous, then the control Hamiltonian space may be described by a two dimensional surface. But the process of control takes place as a temporal sequence and any variation of these parameters must also be a temporal sequence. If we try to map a two-dimensional surface by a one dimensional sequence, we run into the concept of space-filling curves.

There is another way space-filling curves can enter quantum chemical systems. If the state space is continuous and described by two parameters, like the parameters of the control Hamiltonian. Then too we have a surface of states. In certain cases it may also be possible to relate a two-dimensional state space with a two parameter control space, i.e. two relate the two sets of parameters, So that by changing the control space parameters we achieve an equivalent change in the state space. Evolution of the system in state space can be ill-defined when we try to view the process in a continuous way, but using the concept of Bell's "beables" Rabitz et al [1] have shown that a satisfactory picture can be formed in a statistical sense. Then the concept of SFC can apply to the state space too in terms of "beables".

In this work we shall first show that in the mathematical sense a true SFC must a singular object in terms of the control parameters. However, its discrete finite versions do not pose the problems of the singularity. We shall then consider a comparison of different ways of scanning the parameter space, including a finite representation of the Peano-Hilbert curve.

## II. SFC AND SINGULAR PARAMETERS

A SFC must have a Hausdorff (fractal) dimension more than one, otherwise it cannot fill a two-dimensional space even partly. Let us take a small segment of the curve

$$f(x, y) = c \qquad (1)$$

where x, y are the (control) parameters.
So anywhere on the curve the tangent is given by

$$dy/dx = -[\partial f / \partial x]/[\partial f / \partial y] \qquad (2)$$

Now if we consider only a small section of the curve

$$ds = dx\sqrt{1 + (dy/dx)^2} \qquad (3)$$

However, for a fractal of dimension $n$

$$ds = ds_0 \, (\Lambda_0 / \Lambda)^n \, \Lambda \qquad (4)$$

So this goes to infinity as scale parameter ? goes to zero.
Hence dy/dx must be singular at every point of the curve.
Hence either ?f/?y = 0 or ?f/?x is singular at every point. The first option makes the curve trivial. The second option simply states that physically it is impossible to span a two-dimensional parameter space completely using only a single parameter [e.g. time] which remains within a finite, and hence practically realizable range.

Hence a practical control experiment involving a two-parameter space must involve discretization of the space, such as choosing a lattice [ in parameter space]. The experiment then would correspond to a finite number of steps through the lattice, and any particular sequence will correspond to a specific trajectory, and the choice of trajectory should be such as to optimize the output expected or desired.

## III. CONTROL PARAMETER SPACE AND SFC APPROXIMATIONS

In quantum control processes the effect of a sequence of laser pulses needed to achieve any given program of operations has been studied by Ramakrishna, Rabitz et al. [2] They considered a one-dimensional model with a one-dimensional amplitude. In general a randomly polarized signal, or one with a given polarization will have two components of the electrical vector, so the field space would be two dimensional. We can proceed in a similar fashion and obtain formulas which are generalizations involving both components. The unitary operator which gives the evolution of the system can now be described by similar unitary operators

$$U_I = \exp[i\Gamma] V_k \ldots V_1 \tag{5}$$

with $k$ being the pulse number in the sequence,

$$V_k = \exp[C_k \begin{pmatrix} 0 & \sin f \\ -\cos f & 0 \end{pmatrix}] \tag{6}$$

in the two-dimensional vector space formed by $|m\rangle$ and $|m+1\rangle$
and

$$C_k = \int_{t_k}^{t_{k+1}} dt\, \vec{E}.\vec{m}$$

**E** and the dipole **μ** being two-dimensional vectors.

Now, with an unknown orientation of the dipole vector ***μ*** it may be necessary to try out different components of ***E*** to get the optimal $C_k$ and hence it may be necessary to find the best algorithm to hit on the right value. If it is (the discrete practical limit of ) a SFC it may offer certain advantages over a random choice or a linear scan, as we shall see later with a particular numerical example. The point to remember is that on a SFC two dimensional neighborhood values [ with small variations of both components] are tried in sequence before moving out of a region, and physically that may offer the possibility of locating the right $C_k$ quicker than varying one parameter at a time as in a linear scan, because such a trajectory is likely to have a better autocorrelation [ and hence a smoother variation] than any other option.

We may also be interested in a system with an arbitrarily large number of neighboring states each attainable by small changes in control parameters. For example a simple Bohr atom has energy levels going like $\sim 1/n^2$, where $n$ is the principal quantum number. So if we want to jump from any state $|n_1\rangle$ to another $|n_2\rangle$, we need a frequency
$?(n_1, n_2) \sim 1/n_2^2 - 1/n_1^2$
and this will be nondegenerate for different $n_1$ and $n_2$, but almost continuous for $n_1$ and/or $n_2$ large.

Now supposing we need a particular sequence of jumps $(n_1 ? n_2)$ in an ensemble of such atoms, we would need to use $?(n_1, n_2)$ in the given sequence, i.e. we would need to trace out a trajectory in the $(1/n_1^2, 1/n_2^2)$ space (we use the reciprocals to emphasize the approximation to a continuum).

There is a solved problem in mathematics, called the ETSP [ Euclidean traveling salesman problem] which states that the optimum joining a very large random sequence of cities within a disc is given in the infinite limit by the SFC called MNPeano [3] , which is similar to the SFC Peano curve [ Figure 1]. The path length is given by [ in the units for A below]

$$L_{optimum} = 0.72 \sqrt{N A} \qquad (7)$$

where $N$ is the number of states to be excited and $A$ is the area of the control parameter space. This expression is similar to a random walk problem, except for the coefficient, which is somewhat smaller than 1.

So the quickest way to make a large sequence of quantum jumps for an ensemble of Bohr atoms would be given by the discrete approximation of such a MNPeano curve in the control parameter space which would need to be somehow translated into the corresponding frequency of the exciter. Similar situations of greater practical utility may be found involving electron energy bands of materials and continuous spectra of molecules. The former may be relevant to electronic devices.

## IV. MAGNETIC ENVIRONMENT : A NUMERICAL EXAMPLE

NMR spectroscopy is a very useful tool in augmenting our knowledge of the structure of large molecules when X-ray diffraction data or analyses are insufficient. The magnetic environment of a nucleus with nonzero magnetic moment, such as a proton, or $P^{31}$ may be affected by the electron orbits in the neighborhood contributing their own magnetic fields. As a result nuclear magnetic resonance would occur for a shifted field compared to the lone nucleus, which may be expressed as a chemical shift in the Lande $g$ factor.

$$H_I = - g \vec{s} \cdot \vec{H} \mathbf{m}_N \qquad (8)$$

where **s** is the nuclear spin, **H** is the magnetic field at the nucleus, and $\mu_N$ is the nuclear magneton. In nuclear magnetic imaging used in medical diagnosis one displaces a component of the magnetic field spatially as a raster scan and observes the absorption of the electromagnetic waves for the resonant change of spin state.

In theory it is possible to combine these two uses of nuclear resonance to get a spatial picture of the magnetic environment in a given space of the molecule, e.g. the active site of an enzyme. This is apparently a more direct and theoretically more convenient way, involving no ill-posed inversion problem, than another novel method using the Ehrenfest theorem and a classical correspondence proposed by Rabitz and Vivie-Riedle [4].

Let us take a crystal of the molecule and inundate it slow neutrons. Let us use two orthogonal narrow strips [ one along $x$ and the other along $y$ say] of movable magnetic fields in the *z*-direction [ Figure 2] such that their combined value, which occurs at the intersection of the two strips and hence corresponds to a unique *(x, y)* point is near the resonance value for the r.f. radiation used for the neutron. Then if the r.f. freq is varied over a relevant range for each intersection of *(x, y)* and the absorbed frequency observed, we get the magnetic field at that point. This may allow us to draw a magnetic map inside the molecule and hence can give us useful ideas about its structure. The problem we ask now is what should be the trajectory of the control parameters *(x, y)* for an optimal scan. Again, it would appear that the trajectory with the highest autocorrelation would be the best, because it would identify the two-dimensional region in space with smoothness and hence reliability of measurement.

We performed a simulation with four dipoles, all with equal x, y components at four corners of a square, with the field measured in a smaller [with half the sides] square inside the larger one. The field is of course

$$\vec{H} = \sum_i [\mathbf{m}/(4\mathbf{p})][3\vec{m}_i \cdot \vec{r}_i \vec{r}_i / r_i^2 - m_i] / r_i^3 \qquad (9)$$

where $r_i$ is the unit vector from the magnetic moment vector $m_i$ to the observation point and $r_i$ is the magnitude of the distance vector etc.

We calculated the autocorrelations for the *i*-th value in the sequence with the *(i+k)*-th value, both for $H_x$ and $H_y$, using a linear scan and then using a Hilbert-Peano scan, which is a SFC [ finite approximation at a given scale, in this case with 32 points each way]. Table I gives the autocorrelations for $H_x$ ($H_y$ is similar). It is obvious that the Hilbert-Peano trajectory gives the much better correlated sequence than a linear scan.

## V. CONCLUSIONS

We have argued in general terms about the relevance of SFCs in choosing trajectories of control parameters, and have also shown by a simple simulation example that a SFC is better autocorrelated and hence smoother when magnetic data are sequenced. This is similar to the finding in image processing [5] but much more pronounced. More realistic examples and details are being investigated. Quantum control has previously [2] been identified as useful in three contexts: (a) transferring population from an initial state to a given state, (b) making transitions through a given sequence of states, and (c) forcing the system to have a given expectation value for an observable. We add here the possibility of also using quantum control not to make transitions ( the resonant nucleon returns to ground) but to probe the system as it is.

## VI.

The author would like to thank H. Rabitz for useful discussions.

# TABLE I

**Autocorrelation $<H_x(i)\ H_x(i+k)>$**

| k | Linear scan | Hilbert-Peano |
|---|---|---|
| 0 | 1.00 | 1.000 |
| 1 | 0.86 | 0.995 |
| 2 | 0.73 | 0.990 |
| 3 | 0.60 | 0.984 |
| 4 | 0.49 | 0.980 |
| 5 | 0.38 | 0.973 |
| 6 | 0.29 | 0.967 |
| 7 | 0.19 | 0.960 |
| 8 | 0.11 | 0.955 |
| 9 | 0.04 | 0.949 |
| 10 | -0.01 | 0.943 |

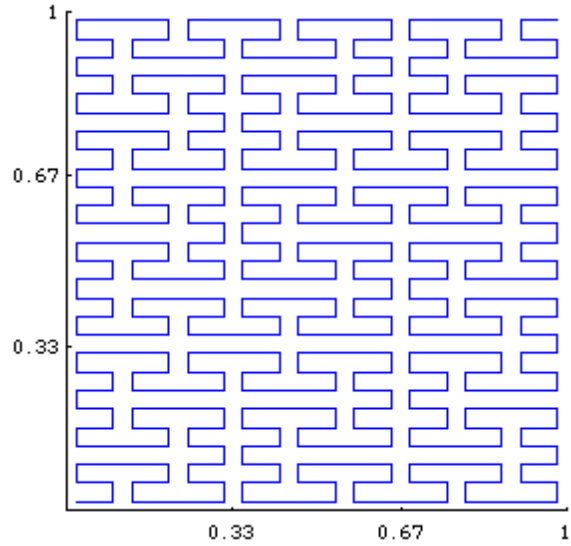

Figure 1: Peano-Hilbert curve [ finite approximation]

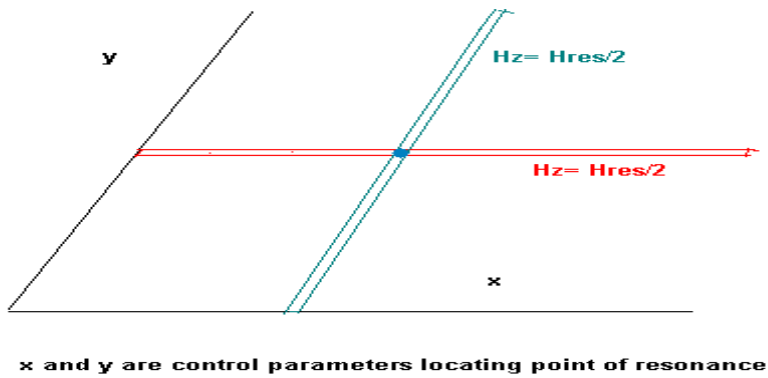

Figure 2: using two orthogonal fields as control to locate point of NM resonance.